\documentclass[pra,graphics,a4paper,10pt,twocolumn,nofootinbib, superscriptaddress,longbibliography]{revtex4-1}

\usepackage[margin=20mm]{geometry}
\usepackage{color}
\usepackage{amssymb,amsmath,amstext}
\usepackage{amsfonts}
\usepackage{braket}
\usepackage{pgfplots}
\usepackage{ mathrsfs }
\usepackage{dsfont}
\usepackage[colorlinks=true,citecolor=blue,linkcolor=magenta]{hyperref}

\newcommand{\tr}{{\rm Tr}}
\renewcommand{\vec}[1]{\boldsymbol{#1}}  

\pgfplotsset{compat=1.16}
\begin{document}

\title{Unified approach to data-driven quantum error mitigation}

\author{Angus Lowe}
\thanks{The first two authors contributed equally to this work.}
\affiliation{Department of Combinatorics and Optimization and Institute for Quantum Computing , University of Waterloo, Waterloo, Ontario N2L 3G1, Canada}
\author{Max Hunter Gordon}
\thanks{The first two authors contributed equally to this work.}
\affiliation{Instituto de F\'{\i}sica Te\'{o}rica, UAM/CSIC, Universidad Aut\'{o}noma de Madrid, Madrid, Spain}
\author{Piotr Czarnik}
\affiliation{Theoretical Division, Los Alamos National Laboratory, Los Alamos, NM 87545, USA}
\author{Andrew Arrasmith}
\affiliation{Theoretical Division, Los Alamos National Laboratory, Los Alamos, NM 87545, USA}
\author{Patrick J. Coles}
\affiliation{Theoretical Division, Los Alamos National Laboratory, Los Alamos, NM 87545, USA}
\affiliation{Quantum Science Center, Oak Ridge, TN 37931, USA}
\author{Lukasz Cincio}
\affiliation{Theoretical Division, Los Alamos National Laboratory, Los Alamos, NM 87545, USA}
\affiliation{Quantum Science Center, Oak Ridge, TN 37931, USA}

\begin{abstract}
Achieving near-term quantum advantage will require effective methods for mitigating hardware noise. Data-driven approaches to error mitigation are promising, with popular examples including zero-noise extrapolation (ZNE) and Clifford data regression (CDR). Here we propose a novel, scalable error mitigation method that conceptually unifies ZNE and CDR. Our approach, called variable-noise Clifford data regression (vnCDR), significantly outperforms these individual methods in numerical benchmarks. vnCDR generates training data first via near-Clifford circuits (which are classically simulable) and second by varying the noise levels in these circuits. We employ a noise model obtained from IBM's Ourense quantum computer to benchmark our method. For the problem of estimating the energy of an 8-qubit Ising model system, vnCDR improves the absolute energy error by a factor of 33 over the unmitigated results and by factors 20 and 1.8 over ZNE and CDR, respectively. For the problem of correcting observables from random quantum circuits with 64 qubits, vnCDR improves the error by factors of 2.7 and 1.5 over ZNE and CDR, respectively. 
\end{abstract} 

\maketitle

\section{Introduction}

Quantum computers are approaching the important milestone of having a demonstrable advantage over classical computers for practical applications, such as chemistry and materials science~\cite{bauer2020quantum}. Such a quantum advantage is expected to be demonstrated with near-term devices that do not have the number of qubits or the gate fidelities required to implement full quantum error correction~\cite{preskill2018quantum}. Nevertheless, the noise of such devices remains a serious obstacle to practical applications~\cite{wang2020noise}. While near-term devices will not be able to completely remove errors caused by device noise, it is often possible to mitigate them.

Such so-called error mitigation (EM) techniques are sure to be an essential part of demonstrating the utility of quantum technologies, for example, for achieving chemical accuracy in chemistry applications. To this end, many distinct EM methods have been proposed~\cite{Temme_2017,endo2018practical}. One approach is to optimize quantum circuits using compiling and machine learning \cite{Cincio_2018, Murali, cincio2020machine}, while another employs variational quantum algorithms~\cite{VQE,mcclean2016theory,cerezo2020variationalreview} to reduce circuit depth and potentially remove the effects of incoherent noise~\cite{Li,QAQC,sharma2019noise,larose2019variational,cirstoiu2020variational,cerezo2020variational}. More recently, quantum phase estimation has been employed for error mitigation~\cite{o2020error}.
\begin{figure}[ht!]
    \centering
    \includegraphics[width=0.49\textwidth]{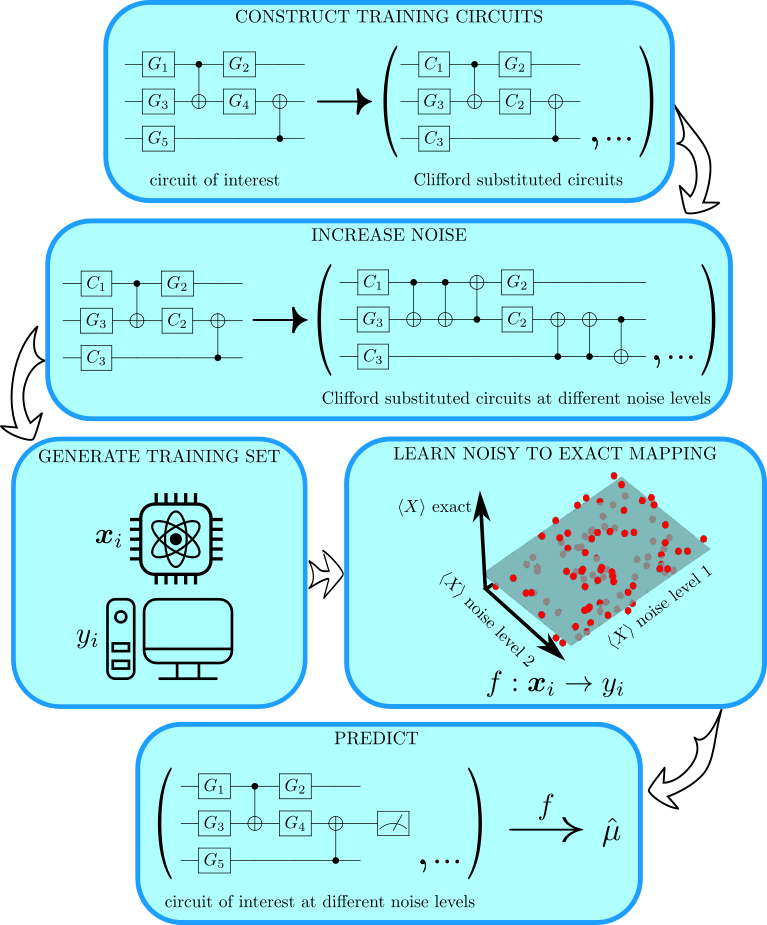}
    \caption{\textbf{The Variable Noise Clifford Data Regression (vnCDR) method.} The first step constructs a set of near-Clifford training circuits that are close, in some sense, to the circuit of interest. The second step increases the size of the training set by adding variable amounts of noise to the circuits generated in the first step. The third step involves both classical simulation and quantum evaluation of the training circuits to generate the noise-free and noisy training data, respectively. The fourth step trains the parameters of an ansatz, which we take as a hyperplane, to fit the training data. Finally, one uses this fitted ansatz to predict the desired observable for the circuit of interest.}
    \label{fig:vnCDR_diagram}
\end{figure}

Zero-noise extrapolation (ZNE) is a classical post-processing approach to EM that has received a significant amount of attention \cite{Temme_2017}. ZNE combines observables evaluated at several controlled noise levels through stretching gate times or inserting identities \cite{Dumitrescu_2018, He_2020, otten2019recovering, giurgica2020digital, cai2020multiexponential}, enabling extrapolation to the zero-noise limit. Despite much success \cite{Kandala_2019}, this method is not without its limitations. Due to the uncertainty of the extrapolation, performance guarantees are difficult in general. In particular, ZNE struggles when a low degree polynomial fit to the noisy expectation values fails to match the behavior in the zero-noise limit. For simple noise models or very low depth circuits this extrapolation can be well behaved. But in real devices using less trivial circuits, the lowest error points available are often too noisy for such fits to be helpful.

Recently, alternative mitigation methods have been developed that make use of learning from data sets constructed using Clifford quantum circuit data \cite{strikis2020learning,czarnik2020error}. These methods are attractive based on their relative simplicity and scalability due to the classically simulable nature of quantum circuits comprised mainly of Clifford gates (gates that map Pauli operators to Pauli operators). 

For example, the Clifford Data Regression (CDR) method~\cite{czarnik2020error} first chooses a training set of near-Clifford quantum circuits related to the circuit of interest. A scalable classical simulator of near-Clifford circuits and a noisy quantum computer are used to compute the noise-free and noisy data, respectively. Finally, the trained ansatz is used to predict the noise-free observable for the quantum circuit of interest.

Both ZNE and CDR are data-driven approaches to error mitigation, but they use different types of data. ZNE uses variable noise data while CDR uses variable Clifford circuit data. A natural question is whether combining these approaches could lead to a unified technique that is more powerful than the individual ones. In this work, we propose a novel method that answers this in the affirmative.

Our approach is called variable noise Clifford data regression (vnCDR). vnCDR considers a collection of near-Clifford training circuits like CDR, each evaluated at multiple noise levels as in ZNE. One can think of this process as either informing the extrapolation in ZNE about the zero-noise limit for similar circuits or as adding relevant features to the regression model in CDR. In the latter view, this is philosophically similar to data augmentation techniques that introduce artificial noise in machine learning~\cite{Goodfellow-et-al-2016}. The ansatz employed in vnCDR is motivated by Richardson extrapolation and by noting it perfectly removes the effects of global depolarizing noise (see Appendix~\ref{app:global_depolarizing}). We also comment that training on the set of Clifford circuits is sufficient as the Clifford gates span the space of single qubit unitaries (see Appendix~\ref{app:sufficiency_of_cliffords}). Figure~\ref{fig:vnCDR_diagram} gives a schematic illustration of vnCDR.

Below we first provide background on ZNE and CDR, and then we present our unified method. Using a noisy simulator based on a gate set tomography of IBM's Ourense quantum computer, we compare the performance of ZNE, CDR, and vnCDR for two tasks. The first task is estimating the energy of an $8$-qubit transverse Ising model with the Quantum Alternating Operator Ansatz (QAOA). Correcting circuits of this form is relevant for both combinatorial optimization problems and condensed matter studies~\cite{qaoa2014,hadfield2019quantum}. Our second task involves random quantum circuits for large qubit numbers (up to $64$ qubits with $6$ CNOT layers) and large circuit depth (up to $16$ CNOT layers with $8$ qubits). The lack of structure in these random circuits makes them a difficult use case for these EM methods, and they give us a notion of these methods' utility in more general settings.

For both use cases, vnCDR outperforms ZNE and CDR. For the QAOA task we analyze the absolute energy error and obtain with vnCDR a factor of $20$ improvement over ZNE and a factor of $1.8$ improvement over CDR. For the random circuit task we obtain in the case of $64$ qubits factors of $2.7$ and $1.5$ improvement over ZNE and CDR, respectively, while for the case of $16$ layers we obtain factors of $2.3$ and $1.3$ improvement over those methods.

\section{Background}

\subsection{Zero-noise extrapolation}\label{sec:zne}

ZNE~\cite{Temme_2017} involves varying the noise level of a quantum circuit to infer the noise-free behavior. Assuming a dependence on noise parameter $\epsilon$, the correction is performed by taking linear combinations of the noisy expectation values in such a way that errors attributable to terms of order $n$ or less are canceled, where $n$ is the number of additional noise levels employed. 

Following the presentation in Ref.~\cite{Temme_2017}, denote the noise-free expectation by $\mu$ and consider the task of correcting the expectation value obtained from a noisy quantum device with noise characterized by parameter $\epsilon$. First, one chooses a set of noise levels $\mathcal{C}=\{c_0,c_1,\dots,c_n|c_0=1, c_j<c_{j+1}\}$ 
and runs the device with amplified noise $c_j\epsilon$ to obtain an estimate $\hat{\mu}_j$ for all noise levels $c_j\in \mathcal{C}$. The final correction $\hat{\mu}$ can then be computed as
\begin{align}\label{eq:zne_eqn}
    \hat{\mu} = \sum_{j=0}^n \gamma_j \hat{\mu}_j
\end{align}
where the set of coefficients $\{\gamma_j\}$ are chosen to satisfy
\begin{align}\label{eq:richardson_extrap}
     \sum_{j=0}^n \gamma_j = 1, \quad \sum_{j=0}^n \gamma_j c_j^k=0 \ \ \forall k\in\{1,\dots,n\}.
\end{align}
This technique, known as Richardson extrapolation~\cite{richardson1927, Temme_2017}, ensures the error of the final estimate is of the order $O(\epsilon^{n+1})$.

Though ZNE was originally proposed in a context where one can stretch gate times to achieve the various noise levels $c_j$, recent work has suggested a hardware agnostic implementation based on identity insertions~\cite{He_2020, Dumitrescu_2018}. For example, inserting 2 CNOT gates applied one after the other is an identity matrix in the noise-free circuit evaluation, but is likely to affect the output in the noisy case. In the fixed identity insertion method (FIIM), the noise levels are taken to be the number of additional gates added in this manner, so inserting 2 additional CNOT gates for every CNOT in the original circuit results in noise levels $c_j = 1,3,5,\dots$ being implemented.

As noted in Ref.~\cite{He_2020}, Richardson extrapolation is equivalent to performing a polynomial interpolation on the various noisy expectations, treating the noise levels $c_j$ as the independent variable. To see this, note that for any solution $\{b_k\}_{k=0}^n$ to the system of equations
\begin{align}\label{eq:poly_fit_eqn}
    \hat{\mu}_j = b_0 + \sum_{k=1}^nb_k c_j^k, \quad j=0,1,\dots,n
\end{align}
it holds that
\begin{align}
    \hat{\mu} &= \sum_{j=0}^n\gamma_j \left(b_0 + \sum_{k=1}^n b_k c_j^k\right)\nonumber\\
    &= b_0 \left(\sum_{j=0}^n \gamma_j\right) + \sum_{k=1}^n b_k\left(\sum_{j=0}^n\gamma_j c_j^k\right)\nonumber\\
    &= b_0.
\end{align}
A unique solution exists to~\eqref{eq:poly_fit_eqn} for $n+1$ distinct noise levels $c_j$ when performing an order-$n$ interpolation. (See for example Ref.~\cite{szego1975orthogonal}.) Alternatively, one can adopt a lower degree polynomial fit as done in our numerical experiments by computing the least-squares solution to the resulting system of equations. For example, if one wishes to perform a linear fit on the data from $n+1$ distinct noise levels, one may write
\begin{align}
    \vec{\mu} = \begin{bmatrix}
    \mu_0\\
    \mu_1\\
    \vdots\\
    \mu_n
    \end{bmatrix},\quad
    \vec{X} = \begin{bmatrix}
    1 & c_0\\
    1 & c_1\\
    \vdots & \vdots\\
    1 & c_n
    \end{bmatrix},\quad
    \vec{b} = \begin{bmatrix}
    b_0\\
    b_1
    \end{bmatrix}
\end{align}
and express the system of equations as
\begin{align}
    \vec{\mu} = \vec{X}\vec{b}.
\end{align}
Taking the y-intercept $b_0$ of the least-squares solution,
\begin{align}
    \vec{b} = \left(\vec{X}^T \vec{X}\right)^{-1}\vec{X}^T \vec{\mu}
\end{align}
yields the extrapolated expectation value.

Crucially, the value of the correction is completely determined by a fixed set of noisy expectations for any choice of the noise amplification and extrapolation techniques above. In particular, Eq.~\eqref{eq:zne_eqn} enforces an $n^\text{th}$ degree polynomial fit when one has data points at $n+1$ noise levels, which may not provide a good approximation to the behavior near $\epsilon=0$ when the data points that are experimentally accessible all reflect a fairly high amount of noise. While some authors have also successfully used lower degree polynomial fits, there is evidence to suggest the resulting corrections can still be fairly inaccurate~\cite{endo2018practical, Dumitrescu_2018,czarnik2020error}.
This motivates our proposal for a method based on learning from efficiently simulable circuits, avoiding some of the drawbacks of ZNE.

\subsection{Clifford data regression}

In CDR \cite{czarnik2020error} the expectation values obtained from a quantum device are corrected using a straightforward linear regression based on examples from circuits comprised mainly of Clifford gates. These Clifford circuits are efficiently simulable and  generated in such a way as to remain similar to the original circuit of interest. Explicitly, the goal is to learn a function which takes noisy expectations to their error mitigated values:
\begin{align}\label{eqn:cdr_ansatz}
    f(\hat{\mu}_0) = a_1\hat{\mu}_0 + a_2
\end{align}
where $\hat{\mu}_0$ is the noisy expectation and the $a_1, a_2$ are parameters chosen optimally by least-squares regression on the Clifford circuit dataset i.e., for a training set of $m$ noisy Clifford circuit expectations $\{x_i\}$ and corresponding targets $\{y_i\}$ obtained via classical simulation, one computes

\begin{align}
    (a_1,a_2) = \underset{(a_1,a_2)}{\text{argmin}} \sum_{i=1}^m \left[y_i - (a_1 x_i + a_2)\right]^2. 
\end{align}

The form of the ansatz can be physically motivated using a simplified noise model. Let $\rho$ be the density matrix for the state of a device which has undergone some noise-free evolution and
consider a global depolarizing noise channel $\mathcal{E}$ which acts on this state before a measurement of the observable $X$. It then holds that
\begin{align}
    \tr(\mathcal{E}(\rho) X) = (1-\epsilon)\tr(\rho X) + \frac{\epsilon\tr(X)}{d}
\label{eqn:depolarizing}
\end{align}
where $d$ is the dimension of the system and $\epsilon$ is a parameter characterizing the noise.  Identifying $\hat{\mu}_0 = \tr(\mathcal{E}(\rho) X)$ and
\begin{align}
    a_1 = 1/(1-\epsilon),\quad a_2 = -\frac{\epsilon}{d(1-\epsilon)}\tr(X)
\end{align}
we see that the desired quantity $\tr(\rho X)$ can be recovered using Eq.~\eqref{eqn:cdr_ansatz}.

When applied to a more realistic noise model obtained from an IBM quantum device, empirical results suggest CDR yields significantly more scalable corrections than those from ZNE~\cite{czarnik2020error}, at least in the plausible setting of being limited to coarse-grained noise amplification. In the following sections, we improve upon the CDR method by incorporating data obtained at variable noise rates, which leads to more accurate predictions of the noise-free expectation values.

\section{The vnCDR Method}

Let $U$ be a quantum circuit, $\ket{0}$ its initial state, and $X$ an observable of interest. Consider the task of estimating the expectation value $\mu = \bra{0}U^\dag X U \ket{0}$ from measurements of a noisy quantum device. The variable noise Clifford data regression (vnCDR) method is performed with the following steps.
\begin{enumerate}  
    \item (Clifford data) Choose a set of circuits $\mathcal{S} = \{V_i\}_{i=1}^m$ based on $U$ which will be used to form the training set $\mathcal{T}$ in step 3. The circuits in $\mathcal{S}$ must be efficient to simulate classically, which is ensured by constructing them primarily from Clifford gates. The number of non-Cliffords used is denoted by $N$. Note that $N$ is assumed to be a constant parameter here, so the simulations are classically tractable.

    \item (Noise data) Choose a set of noise levels $\mathcal{C}=\{c_0,c_1,\dots,c_n\}$ where $1 = c_0 < c_1 < \dots < c_n$ which will be used to form the training set $\mathcal{T}$ in step 3. If the noise is characterized by a parameter $\epsilon$ then running the device with noise level $c_j$ means the new parameter is $c_j\epsilon$.
    
    \item (Training set)  For each of the $m$ circuits $V_i$ in $\mathcal{S}$ and $n+1$ noise levels $c_j\in \mathcal{C}$, produce an estimate of the observable expectation called $x_{i, j}$. Also, for each of the $m$ circuits compute $y_i = \bra{0}V_i^\dag X V_i \ket{0}$ using a classical simulation. The training set $\mathcal{T}$ is then defined as $\mathcal{T} = \{(\vec{x}_i, y_i)\}$ where $\vec{x}_i = (x_{i,0},\dots,x_{i,n})$ is the vector of noisy estimates originating from the $i^\text{th}$ circuit.

    \item (Learning) Learn a function $f : \mathbb{R}^n \to \mathbb{R}$ that takes a set of noisy estimates at the $n+1$ different noise levels and outputs an estimate for the noise-free value. Specifically, we take the linear ansatz $g : \mathbb{R}^n \times \mathbb{R}^n \to \mathbb{R}$,
        \begin{equation}
           g(\vec{x}; \vec{a}) = \vec{a}\cdot\vec{x}\,.
        \end{equation}
    We use least-squares regression on the dataset $\mathcal{T}$ to pick optimal parameters $\vec{a}^*$, i.e.,
    \begin{align}
        \vec{a}^* = \underset{\vec{a}}{\text{argmin}} \sum_{i=1}^m \left(y_i - g(\vec{x}_i; \vec{a})\right)^2\,,
    \end{align}
    so that we expect $f(\vec{x}) = g(\vec{x}; \vec{a}^*)$ to output a good estimate for the noise-free expected value given a vector of noisy ones.
    \item (Correction) Use the estimate $\hat{\mu} = \vec{a}^*\cdot\vec{\mu}$, where $\vec{\mu} = (\hat{\mu}_0,\dots,\hat{\mu}_n)$ is comprised of the $n+1$ noisy expectations for the original circuit.
\end{enumerate}

Our method shares common features with both ZNE and CDR. Specifically, the functional form of the ansatz we choose resembles a Richardson extrapolation on the noisy expected values as shown in Section~\ref{sec:zne}. However, the method differs in its approach to relating the noisy values to the final estimate. Namely, in ZNE, the output is a fixed function of the various $\hat{\mu}_j$, whereas vnCDR attempts to learn the best candidate from a family of functions parametrized by $\vec{a}$. In a certain sense, vnCDR is thus choosing the best possible extrapolation of the noisy data from the original circuit using examples from Clifford circuits which are similar in structure. 

The method can also be viewed as adding relevant variables to the CDR method, before performing a multiple linear regression on the new dataset. This description comes with the caveat that -- in contrast with the CDR ansatz (see Eq.~\eqref{eqn:cdr_ansatz}) -- the new parametrization $g(\vec{x};\vec{a})$ is a linear mapping without a constant term. There are two motivations for this. Firstly, such a parametrization corresponds well with the linear combination of noisy expectations that is utilized in the ZNE method (Equation~\eqref{eq:zne_eqn}). Secondly, restricting the class of functions we are searching over to be linear awards us an intuitively desirable property: if the function we arrive at achieves zero error on all circuits composed of Clifford gates, then it will predict the expectation values of arbitrary circuits with zero error. This result boils down to the observation that Clifford gates span the space of single-qubit unitaries. (See Appendix~\ref{app:sufficiency_of_cliffords} for the proof of this statement.)

The form of the ansatz can be further motivated by considering the action of a global depolarizing channel (see Eq.~\eqref{eqn:depolarizing}). We note that this simple model was proposed recently to effectively describe dominant effects of the noise in real devices~\cite{vovrosh2021efficient,urbanek2021mitigating}.   The vnCDR ansatz can be shown to completely mitigate the effect of such a channel on some observable of interest (see Appendix~\ref{app:global_depolarizing}), similar to CDR. 
\section{Numerical results}
\label{sec:results}
\subsection{Transverse-field Ising model}
\label{sec:results_qaoa}
\begin{figure}[t!]
    \centering
    \includegraphics[width=0.45\textwidth]{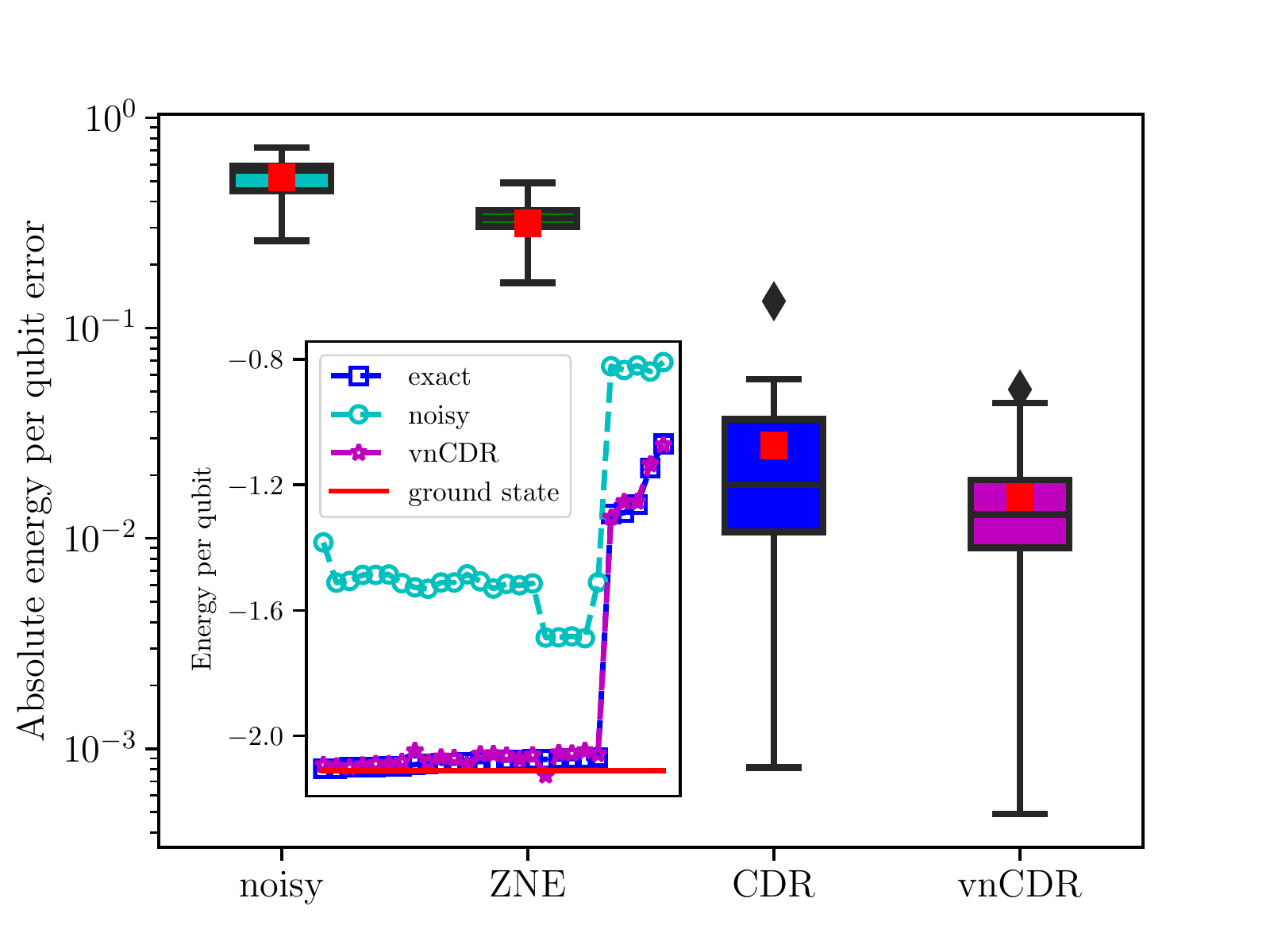}
    \caption{\label{fig:vncdr_ising_q8} Correcting 27 local minima of the Quantum Ising model (\ref{H})  energy minimization for $Q=8$ qubits. As a variational ansatz we use the QAOA  (\ref{QAOA}) with $p=4$ layers.  We compare absolute errors of the noisy and corrected energies for different error mitigation methods. The results obtained are shown as a box plot where boxes show the interval between the first and the third quartile. The  red squares denote the mean values while the central lines represent the median values.  The whiskers show the range of the data outside the quartiles and diamonds indicate outliers. The inset displays the energies per qubit of the minima calculated by the QAOA energy minimization. }
\end{figure}

First, we consider a task of variational simulation  of the ground state of a 1-D transverse-field Ising model using parameterized quantum circuits.   
The Hamiltonian of the system is given by
\begin{align}
    H = -g \sum_j \sigma_X^j - \sum_{\langle j,j^\prime\rangle}\sigma_Z^j\sigma_Z^{j^\prime},
    \label{H}
\end{align}
where $\sigma_X,\sigma_Z$ are Pauli matrices and $\langle j,j^\prime\rangle$ are nearest neighbor sites on the lattice. We assume here open boundary conditions. We consider the case of $g=2$  corresponding to a paramagnetic phase. We use the QAOA~\cite{qaoa2014,hadfield2019quantum} 
\begin{equation}
\ket{\psi(\vec{\beta},\vec{\gamma})} =  \prod_{j=p,p-1\dots,1} e^{-i \beta_j H_2} e^{-i \gamma_j H_1}  (|+\rangle)^{\otimes Q},
\label{QAOA}
\end{equation}
where $\vec{\beta}$ ,$\vec{\gamma}$ are the rotation angles to be optimized,  $H_1=\sum_{\langle j,j^\prime\rangle}\sigma_Z^j\sigma_Z^{j^\prime}$, $H_2=\sum_j \sigma_X^j$, $\ket{+}=\frac{1}{\sqrt{2}}(\ket{0}+\ket{1})$, and $Q$ is the number of qubits. A decomposition of (\ref{QAOA}) to a quantum circuit is described in Appendix~\ref{app:QAOA_circuit}.

We perform the optimization for $Q=8$ qubits using a circuit depth $p=4$. We minimize the energy evaluated with a noisy simulator using a MATLAB implementation of quasi-Newton gradient descent.  The noise model we employ is obtained by gate set tomography of IBM's Ourense quantum computer and described in detail in Ref.~\onlinecite{cincio2020machine}. Furthermore, we assume perfect measurement as measurement errors can be mitigated by specialized techniques~\cite{chow2010detecting, bravyi2021mitigating}.   To carry out the benchmark of our method, we run 27 instances of the optimization and correct the resulting observable expectations using the ZNE, CDR and vnCDR methods. The corrections are realized on each of the 1- and 2-qubit terms which make up the Hamiltonian from which we then estimate the ground state energy. The results are summarized in Fig.~\ref{fig:vncdr_ising_q8} showing vnCDR outperforms ZNE and CDR with a factor of $33$ improvement of the mean absolute energy error while ZNE and CDR give a factor of $1.7$ and $19$, respectively. 

In the case of  CDR and vnCDR for each of the circuits we construct training sets with $80$ classically simulable circuits, setting the number of non-Cliffords to $N=16$. We remark that there are $60$ non-Clifford gates in total for the circuit of interest. For further information regarding the construction of our training sets, see Section~\ref{sec:implementation_cdr}. 

For the vnCDR and ZNE corrections, we computed the expectation values using the set of noise levels $\mathcal{C} = \{1,3,5\}$ and the fixed identity insertion noise amplification method~\cite{He_2020} which we elaborate upon in Section~\ref{sec:implementation_zne}. The noise level is defined  as the ratio of CNOT gates in the modified circuits compared to the original one. Note that this is a fairly coarse-grained set of noise levels, which may explain why the ZNE performance is quite poor. We also found that including in $\mathcal{C}$ noise levels higher than $5$ did not improve the performance of the methods.
For further details, see Sections~\ref{sec:implementation_zne} and \ref{sec:implementation_vncdr}.

\subsection{Random quantum circuits}
\label{sec:results_rqc}
\begin{figure}
\includegraphics[width=0.48\textwidth]{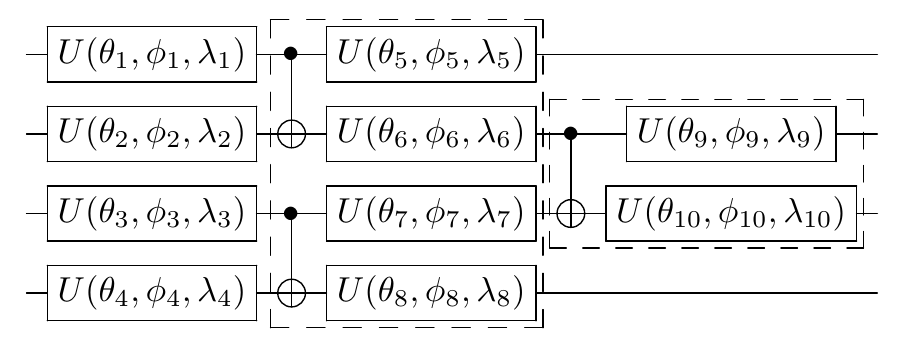}
    \caption{An  example of  the IBMQ hardware efficient ansatz  with $p=2$ layers for $Q=4$ qubits. The layers, represented by gates within the dashed contours, act on a random product state created by general single  qubit unitaries $U$.  The general unitary is defined as  $U(\theta,\phi,\lambda) = R_Z(\phi+\pi)R_X(\pi/2)R_Z(\theta+\pi)R_X(\pi/2)R_Z(\lambda)$, where $R_Z(\alpha) = e^{-i\alpha/2 \sigma_Z}$,  $R_X(\alpha) = e^{-i\alpha/2 \sigma_X}$. Each layer consists of CNOTs interleaved with the $U$ gates.  The CNOT structure alternates between neighboring layers.  We choose angles $\theta$, $\phi$, $\lambda$ of each $U$ gate randomly creating a random quantum circuit. Note that  $R_X(\pi/2)$, CNOTs and $R_Z(\alpha)$ are native gates of the IBM computers. Furthermore, CNOTs and $R_X(\pi/2)$ are Clifford gates.  }
   \label{fig:RQC_example}
\end{figure}

\begin{figure*}[]
    \centering
    \includegraphics[width=0.4\textwidth]{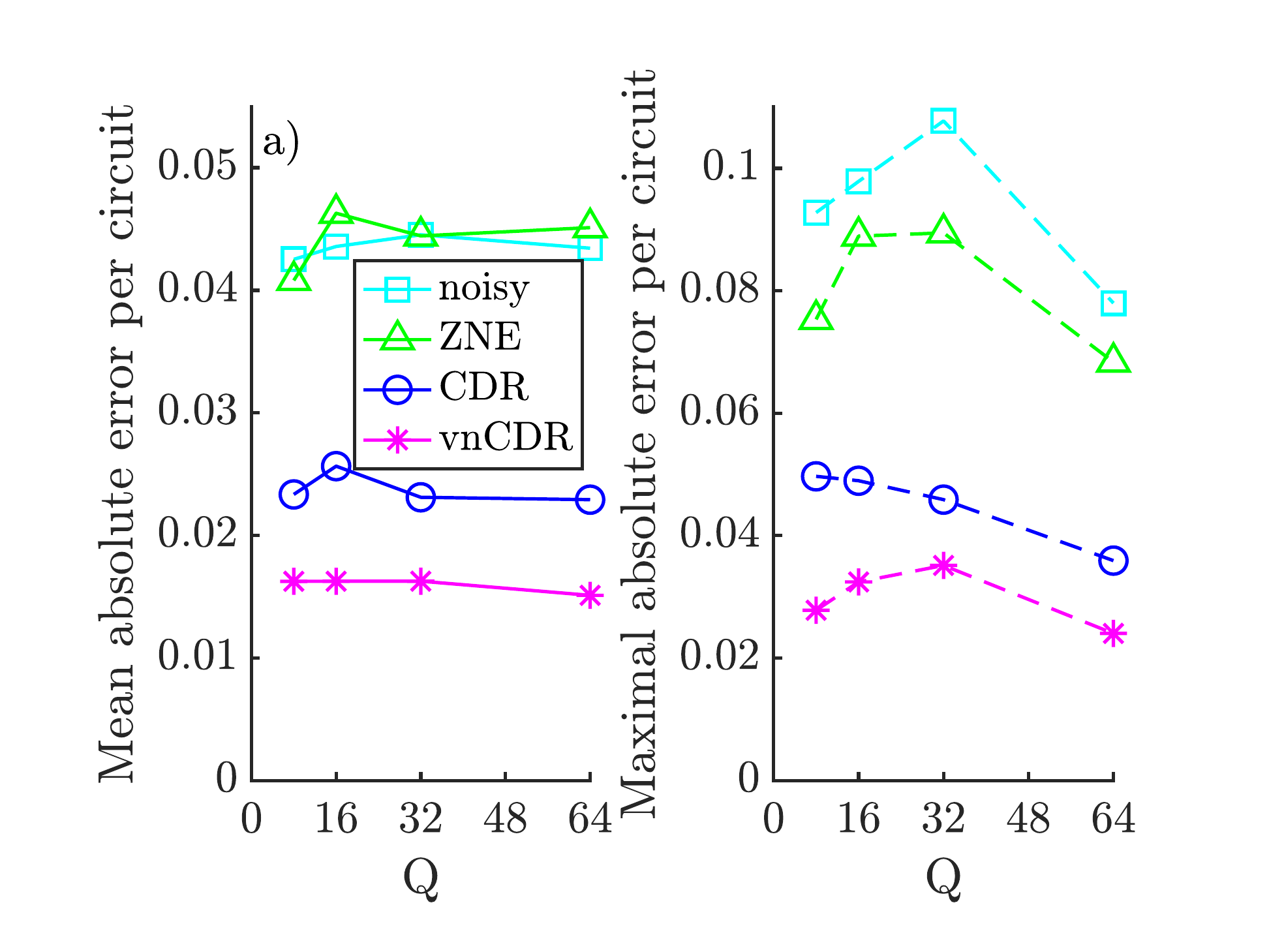}
    \includegraphics[width=0.4\textwidth]{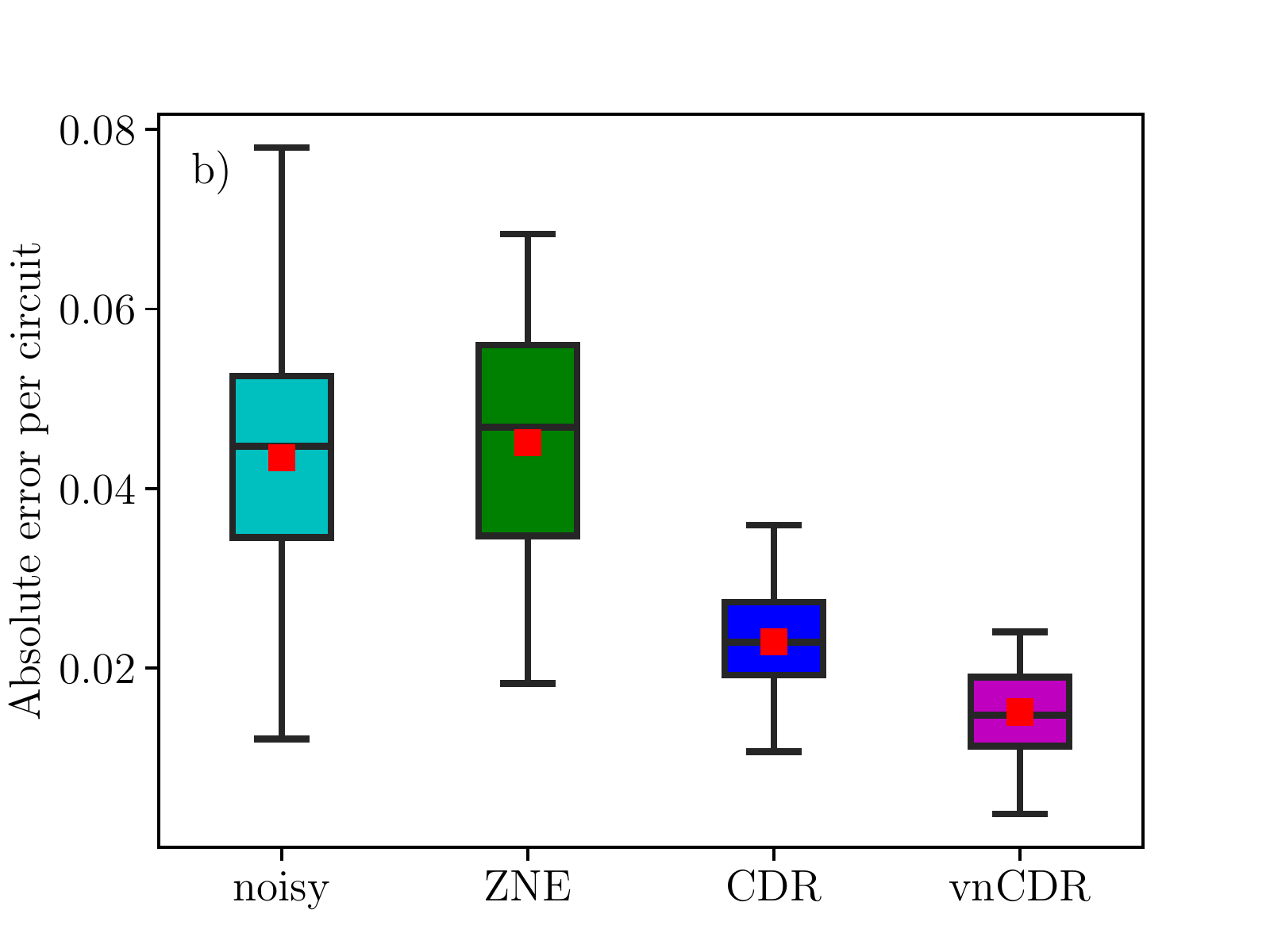}
    \includegraphics[width=0.4\textwidth]{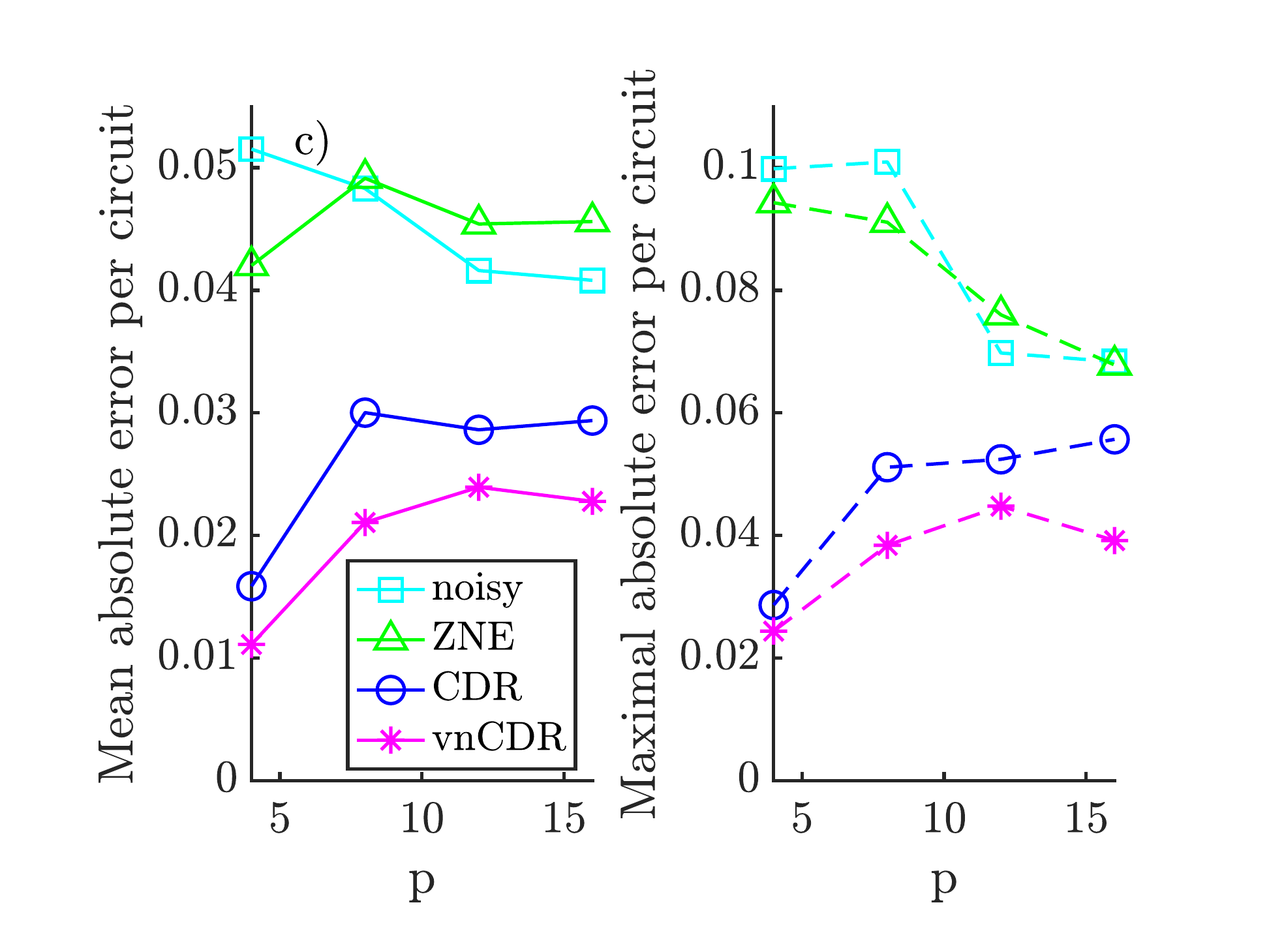}
    \includegraphics[width=0.4\textwidth]{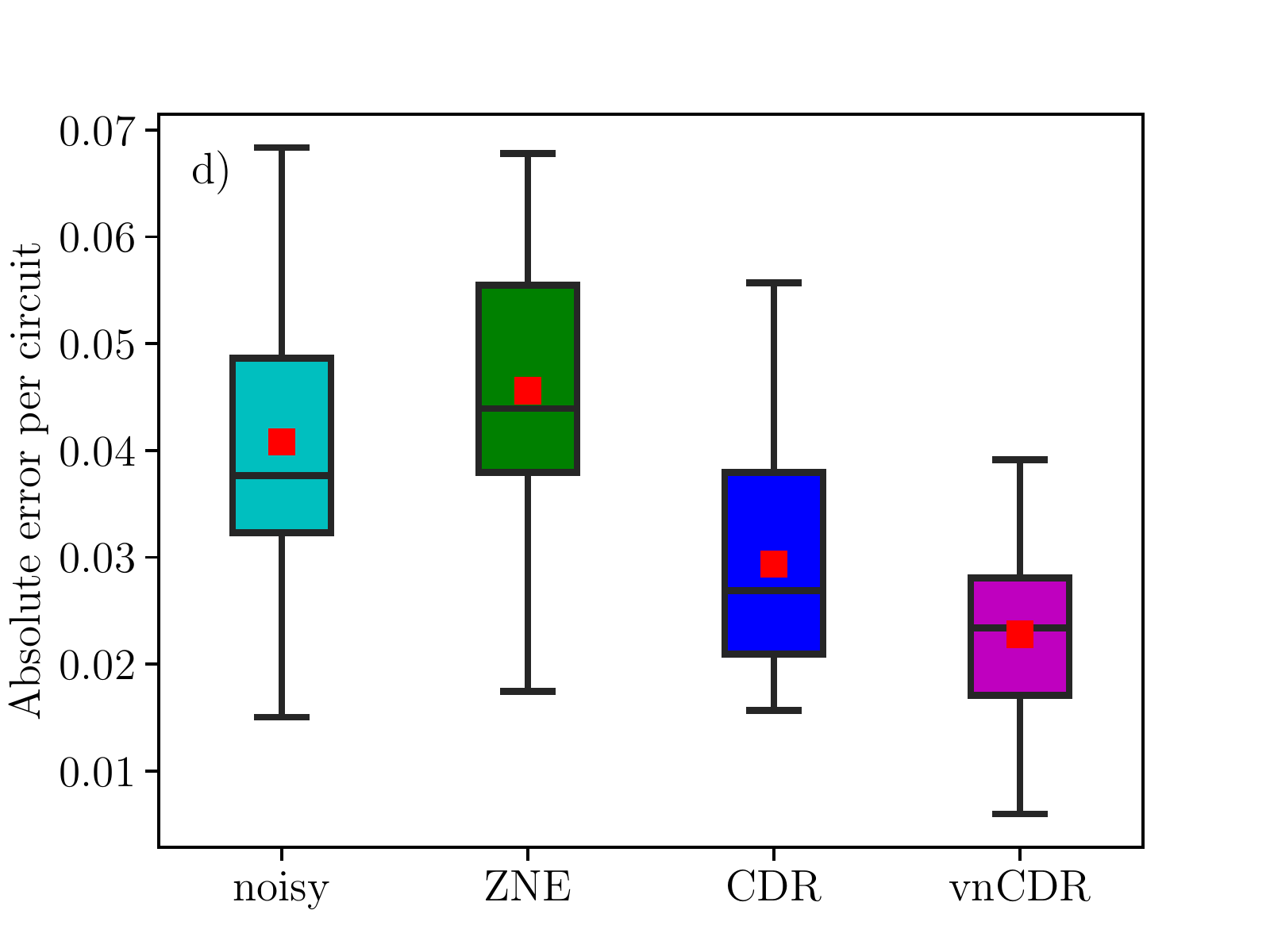}
    \caption{Correcting the IBMQ hardware efficient  ansatz with random parameters. 
    For each value of $Q$ and $p$ we analyze 30 random circuits and  correct 
    $\sigma^{1}_{X}$, $\sigma^{Q/2}_{X}$,  $\sigma^{1}_{Z}\sigma^{2}_{Z}$, $\sigma^{Q/2}_{Z}\sigma^{Q/2+1}_{Z}$ 
    for each of them. We define an absolute error per circuit as the mean of the observables' absolute errors. The absolute  corrected (noisy) observable  error is defined as an absolute value of its difference with respect to the exact value.
    In (a) we  display the scaling with $Q$ of the mitigated and unmitigated absolute error per circuit  for $p=6$. In the left panel we show the mean values  (the solid lines), while the right panel shows the maximal values (the dashed lines).  
     In (b) the bar plot of the error for $Q=64$ and $p=6$.  In (c) the scaling with increasing $p$ for $Q=8$ and in (d) the results for $Q=8$, $p=16$.}
   \label{fig:RQC_scaling}
\end{figure*}

Next we consider an implementation of the IBMQ hardware efficient ansatz with random parameters, see Fig.~\ref{fig:RQC_example}.  The ansatz consists of  layers of  alternating  nearest-neighbor CNOTs  decorated with general one qubit unitaries  $U(\alpha,\beta,\gamma)$. 
We compute one and two qubit observables for $30$ random instances and  correct them with ZNE, CDR and vnCDR methods, see details in the caption of Fig.~\ref{fig:RQC_scaling}. We analyze scaling of the observables absolute error with increasing $Q=8,16,32,64$ for $p=6$ and the scaling with increasing $p=4,8,12,16$  for $Q=8$. To simulate large $Q$ systems we employ a Matrix Product Operators (MPO)~\cite{FannesMPS} noisy simulator with the same noise model as in the case of the Ising QAOA simulations. We discuss the simulator in more detail in Appendix~\ref{app:MPO}. Here to simplify presentation we show results obtained in the limit of infinite shot number. In Appendix~\ref{app:resource_scaling} we show that qualitatively the same  results can be obtained using finite shot numbers feasible with current quantum computers.

The results are discussed in detail in  Fig.~\ref{fig:RQC_scaling}. We find that the vnCDR outperforms ZNE and CDR methods for all simulated $Q$ and $p$ values. For the largest system size, $Q=64$, vnCDR gives a factor $2.7$ improvement of the mean error relative to the noisy results, while ZNE and CDR give factors of $1.0$ and $1.8$. For the deepest $p=16$ the factors are $2.4$, $1.0$ and $1.8$, respectively.

For fixed $p=6$ we observe that the unmitigated mean absolute error does not grow with increasing $Q$ in the limit of large $Q$. Such behavior can be explained by the existence of a threshold $Q$ value for which the causal cones of the observables~\cite{VidalMERA}, stop increasing with $Q$. The causal cone is defined here as gates which affect the expectation value of the observable.  See Appendix~\ref{app:causal_cone} for an example of causal cone construction. We take the causal cone into account when forming vnCDR and CDR training sets, see details in Section~\ref{sec:implementation}. For such an implementation we find that the vnCDR and CDR mitigated mean errors also do not increase with increasing $Q$.
We remark that our noise model does not include cross-talk which in principle may result in a faster increase in the number of gates in the noisy observables causal cones. We leave investigation of the scaling in the presence of such noise to a future work.  
With increasing $p$ we find that quality of the correction decreases for all methods.  Nevertheless,  even in the case of the deepest circuits, $p=16$, we obtain a significant improvement when employing vnCDR.

To perform CDR and vnCDR for each observable of interest in each random circuit we construct a training set using  $100$ classically simulable circuits with  $N=20$ non-Clifford gates.  Detailed discussion of the method used to construct the training sets is given in  Section~\ref{sec:implementation_cdr}. We remark that in the case of $p=6$, $Q=64$ 
circuits the largest number of non-Clifford gates within the causal cone of an observable is $60$,  while for $Q=8$, $p=16$ it  is $312$.

For the vnCDR and ZNE corrections, we increase the noise level by identity insertions as in the case of the QAOA Ising simulations. We find that in both cases it is beneficial to include higher noise levels than in the Ising case:  $\mathcal{C} = \{1,3,5,7,9\}$. We remark that the QAOA circuit having $16$ layers of CNOTs is deeper and than most circuits considered here. As ZNE is supposed to correctly capture noise effects for sufficiently small noise this may explain why it is beneficial to use higher noise levels in the random quantum circuits case.  We leave systematic investigation of this effect to future work.  For more detailed description of the ZNE and vnCDR  implementations see Section~\ref{sec:implementation_zne}, \ref{sec:implementation_vncdr}.

\section{Implementation details}
\label{sec:implementation}
\subsection{ZNE}
\label{sec:implementation_zne}
We perform the noise amplification in our numerical experiments using identity insertions after each application of a CNOT gate~\cite{Dumitrescu_2018,He_2020}. We use the fixed identity insertion method (FIIM) of Ref.~\cite{He_2020}, which adds pairs of CNOT gates after each CNOT gate of the original circuit.   The noise level is defined as the factor by which the number of CNOT gates in the circuit increases. In the first example -- the QAOA optimization task -- we employ noise levels $\mathcal{C} = \{1,3,5\}$, whereas for random circuits we achieved better results with a higher maximum noise level, so we used the set $\mathcal{C} = \{1,3,5,7,9\}$.
We obtained corrected values of the observables of interest by an extrapolation using both a polynomial fit via Eq.~\eqref{eq:zne_eqn} and a linear fit to the data, as explained in Section~\ref{sec:zne}.
In both the Ising and random quantum circuits cases, we found that a linear fit performed better than a polynomial regression for extrapolation, so we report those results here.

\subsection{CDR}
\label{sec:implementation_cdr}
To construct the training set for a circuit of interest we substitute most of the non-Clifford gates in the circuit by Clifford gates with two different substitution strategies, which are explained below. Such a procedure ensures that circuits in the training set are classically simulable and biased towards the circuit of interest. Here we consider circuits of interest  which are compiled for the IBMQ quantum computers. The compiled circuits are built from CNOTs, 
 $R_X(\pi/2)$ pulses and general $\sigma_Z$ rotations $R_Z(\beta)=e^{-i\beta/2 \sigma_Z}$ with $\beta\in[0,2\pi)$. The pulses and CNOTs are Clifford gates while $R_Z(\beta)$ is a Clifford gate only for $\beta = n\pi/2$, where $n$ is an integer. Therefore, we substitute most of the
$R_Z$ gates by $S^n$, where $n=0,1,2,3$ and $S= e^{i\pi/4\sigma_Z}$ is the phase gate. In both the Ising and random quantum circuits cases we find that substantial error reduction can be obtained using training sets built with approximately  $100$  near-Clifford  circuits.

We consider two different substitution strategies.  
The first one   substitutes a randomly chosen non-Clifford $R_Z(\beta)$ by  $S^n$ minimizing  $d(\beta,n) = ||R_{z}(\beta)-S^{n}||$, where $||.||$ is the Frobenius norm.  This procedure is repeated until $N$  non-Clifford  rotations are left in the circuit. We find that this very simple strategy works well for the Ising model enabling us to obtain a factor of $33$ improvement in  calculating the energy. 

In the more general and challenging case of random quantum circuit simulations we find that better results can be obtained with a more sophisticated substitution method. 
 In such a case to construct a training set we tailor our choice of classically simulable circuits to an  observable of interest, substituting  all non-Clifford gates outside its causal cone.  By the causal cone definition  such a replacement does not affect its expectation value. See Appendix~\ref{app:causal_cone} for a discussion of the causal cone construction. Taking into account the causal cone of the observable is especially important in the case of  local observables and large $Q$ shallow circuits because in such a case the causal cone contains only a small fraction of all non-Clifford gates of the circuit of interest. Furthermore,
 for remaining non-Clifford  rotations within the causal cone of the observable of interest we choose both which gate to replace and what gate to replace it with ($S^{n}$) according to a probability distribution 
 $p(\beta_i,n) \propto   e^{-d(\beta_i,n)^{2}/\sigma^{2}}$. Here $i$ numbers the remaining  non-Clifford rotations in the causal cone. We repeat the procedure until $N$  non-Clifford  rotations are left in the  causal cone of the observable of  interest. Here we use $\sigma=0.5$.  Such a choice of the probability distribution  tends to leave gates which would be most severely distorted by the replacement in the circuit, unchanged.  At the same time it produces  more diverse training sets than a direct replacement by the closest power of~$S$.  
We observe that in the case of the random quantum circuits the correction
is more challenging as expectation values of the observable of interest become more clustered around $0$ with increasing $p$. 
Furthermore, we observe that  training sets created by the simple substitution method tend to have exact expectation values clustered around $0$ more strongly than expectation values of the observable of interest. The more sophisticated procedure generates training sets with more diverse exact expectation values. 

\subsection{vnCDR}
\label{sec:implementation_vncdr}
To construct a vnCDR training set we choose the same classically simulable circuits which are used for a CDR training set. We 
also use the same choice of noise levels as used in the ZNE implementation, namely $\mathcal{C} = \{1,3,5\}$ for the Ising and $\mathcal{C} = \{1,3,5,7,9\}$ for the random quantum circuits mitigation. As in the case of ZNE we observe that including more than $5$ noise levels does not improve results for the Ising while it is beneficial for the random quantum circuits case.   
\section{Conclusions}

Data-driven error mitigation involves collecting data from multiple different quantum circuits in order to inform the correction of errors in a particular circuit of interest. In this work, we conceptually unified two distinct, popular methods for data-driven error mitigation: zero-noise extrapolation (ZNE) and Clifford data regression (CDR). Our unified approach, called variable-noise Clifford data regression (vnCDR), appears to be more powerful than the individual methods of ZNE and CDR.

The vnCDR method generates training data from classically simulable near-Clifford circuits, whose noise levels are varied (e.g., by identity insertions). The method then learns how to correct observables on these training circuits. This involves fitting a multi-dimensional ansatz, which we assume is a hyperplane, to the training data.  This enables a guided extrapolation to the noiseless expectation value for the circuit of interest, which dramatically improves the mitigation realized. Rather than doing uninformed extrapolation as in ZNE, the vnCDR method demonstrates that near-Clifford circuits provide an effective guide for the extrapolation process.  The fitted ansatz can be further motivated by considering the effect of a global depolarizing channel on some observable of interest. The effect of such a channel is completely removed using the vnCDR ansatz.

We compared vnCDR to both ZNE and CDR on two tasks: correcting the  energy  of an Ising transverse spin chain and mitigating local observables of random quantum circuits. 
For both of them we used a realistic noise model obtained by gate set tomography of IBM's Ourense quantum computer.   
On each of these tasks, vnCDR outperforms both of these state-of-the-art error mitigation methods. Compared to ZNE, vnCDR was shown to tolerate the relatively high noise levels obtained via fixed identity insertions.

Though preliminary scaling results are promising, further testing on real quantum devices  will help determine the number of non-Clifford gates and size of the training sets required to attain accurate predictions.   It will also help determine limitations of the method while dealing with large and deep noisy circuits which are challenging for  error mitigation methods.  Additionally it would be interesting to apply the vnCDR method using more sophisticated or fine-grained noise amplification schemes such as random identity insertions or pulse stretching. This may enhance performance for the deep circuits necessary to obtain a quantum advantage. In this regime, we envision that vnCDR could play an important role in yielding quantum advantage for chemistry, materials science, and other applications. Finally we note that further testing is necessary to determine the potential of our method for quantum computing architectures with  gate sets other  than IBM's gate set.

\section{Acknowledgements}

This work was supported by the Quantum Science Center (QSC), a National Quantum Information Science Research Center of the U.S. Department of Energy (DOE).  AL and MHG were initially supported by the U.S. DOE through a quantum computing program sponsored by the LANL Information Science \& Technology Institute. Piotr C. and AA were also supported by the Laboratory Directed Research and Development (LDRD) program of Los Alamos National Laboratory (LANL) under project numbers 20190659PRD4 and 20210116DR, respectively. PJC also acknowledges initial support from the LANL ASC Beyond Moore's Law project. LC was also initially supported by the U.S. DOE, Office of Science, Office of Advanced Scientific Computing Research, under the Quantum Computing Application Teams~program.

\bibliography{main.bib}
\appendix
\section{Perfect mitigation of global depolarizing noise }
\label{app:global_depolarizing}
To motivate the form of the vnCDR ansatz we consider the action of a global depolarizing channel (see Eq.~\eqref{eqn:depolarizing}). Assuming this channel acts in our circuit $j$ different times the final state can be written as 
\begin{equation}
    \rho_{j} = (1-\epsilon)^{j}\rho +(1-(1-\epsilon)^{j}) \frac{\mathds{1}}{d}
\end{equation}
where $d$ is  the  dimension  of  the  system  and $\epsilon$ is  a parameter characterizing the noise. Considering the effect of the above channel on $X$ leads to
\begin{align}
    X^{\text{noisy}}_{j} &= \tr(\rho_{j} X) \\
    &=(1-\epsilon)^{j} \mu + (1-(1-\epsilon)^{j})\frac{\tr(X)}{d}
\end{align}
where $\mu=\tr(\rho X)$. As previously discussed the vnCDR ansatz combines evaluations of the observable of interest at various noise levels:
\begin{equation}
    \hat{\mu} = \sum_{j=1}^{n}a^{*}_{j}X^{\text{noisy}}_{j}
\end{equation}
where the parameters $a^{*}_{j}$ are chosen by fitting data produced by near-Clifford circuits. The above expression can be expanded
\begin{equation}
    \hat{\mu} = \sum_{j=1}^{n} a^{*}_{j}\bigg((1-\epsilon)^{j} \mu + (1-(1-\epsilon)^{j})\frac{\tr(X)}{d}\bigg).
\end{equation}
Therefore, for the vnCDR ansatz to completely mitigate the effects of global depolarizing noise, such that $\hat{\mu} = \mu$, we require:
\begin{equation}
    \sum_{j=1}^n a^{*}_{j} (1-\epsilon)^{j} = 1, \quad \sum_{j=1}^n a^{*}_{j}(1-(1-\epsilon)^{j}) = 0,
\end{equation}
or equivalently,
\begin{equation}
        \sum_{j=1}^n a^{*}_{j} = 1, \quad \sum_{j=1}^n a^{*}_{j}(1-(1-\epsilon)^{j}) = 0.
\label{eqn:param_constraints}
\end{equation}
The training circuit observables and the observable of interest will behave the same way under such a noise channel. As such, the fitted parameters $a^{*}_{j}$ will obey the above relations (Eq.~\eqref{eqn:param_constraints}). Therefore, vnCDR can be seen to perfectly mitigate global depolarizing noise for two or more noise levels. 

It is interesting to consider how this contrasts with the ZNE implementation in this work. We used linear extrapolation to the zero noise limit and least-squares fitting of the noisy expectation values for the observable of interest. This extrapolation method is not expected to perfectly mitigate the effect of global depolarizing noise. An exponential extrapolation would be required in order to perfectly mitigate the effects of this channel and polynomial extrapolation is expected to perform better than linear. However, for our simulations we found in general a simple linear extrapolation gave better results.

\section{Sufficiency of the Clifford training set}\label{app:sufficiency_of_cliffords}
Consider a quantum circuit acting on $Q$ qubits which is represented by a noise-free unitary channel $\mathcal{U}$ and let $\rho_0 \in \mathbb{C}^{d\times d}$ be the initial state, where $d=2^Q$. Also suppose we have some observable of interest $X$ and a collection of channels $\mathcal{E}_0, \mathcal{E}_1,\dots, \mathcal{E}_n$ representing $n+1$ different noise levels. Running the circuit with the $j^\text{th}$ noise channel returns the value $\tr\left((\mathcal{E}_j \circ \mathcal{U})(\rho_0)X\right)$ in expectation.

Additionally, for all $j\in\{0,\dots,n\}$ define $\mu_j(\mathcal{V})$ for some unitary channel $\mathcal{V}$ in the following way:
\begin{align}
    \hat{\mu}_j(\mathcal{V}) = \tr\left((\mathcal{E}_j \circ \mathcal{V})(\rho_0)X\right)
\end{align}
and define $\vec{\mu}(\mathcal{V})\in \mathbb{R}^{n+1}$ as
\begin{align}
    \vec{\mu}(\mathcal{V}) = \left( \hat{\mu}_0(\mathcal{V}),\hat{\mu}_1(\mathcal{V}),\dots, \hat{\mu}_n(\mathcal{V}) \right).
\end{align}

For a given circuit $\mathcal{V}$, the vnCDR correction is then given by $f(\vec{\mu}(\mathcal{V})) = \vec{a}\cdot \vec{\mu}(\mathcal{V})$ where $\vec{a}$ is some optimal set of parameters obtained by training the model. Our goal is to show that if the vnCDR estimate $f(\vec{\mu}(\mathcal{C}))$ is fully accurate for all Clifford circuits $\mathcal{C}$, then the output of $f(\vec{\mu}(\mathcal{U}))$ is also accurate for estimating the value $\tr\left(\mathcal{U}(\rho_0)X\right)$.

We begin by observing that the action of a non-Clifford rotation gate can be decomposed into Clifford maps since they span the space of single-qubit unitaries. Therefore, we can write
\begin{align}
    \mathcal{U} = \sum_{j_1}\alpha_{j_1}\mathcal{C}_{j_1}
\end{align}
where $\mathcal{C}_{j_1}$ is the unitary map resulting from replacing one of the non-Clifford rotation gates in the circuit by the $j_1^\text{th}$ Clifford in the basis. Repeating this process recursively for each of the $k$ non-Cliffords in the circuit, we obtain
\begin{align}
    \mathcal{U} = \sum_{j_1,j_2,\dots,j_k}\alpha_{j_1}\alpha_{j_2}\dots\alpha_{j_k} \mathcal{C}_{j_1,\dots,j_k}.
\end{align}
Each of the unitary maps $\mathcal{C}_{j_1,\dots,j_k}$ is now comprised of Clifford maps only. Furthermore, by the linearity of the trace, noise channels, and function $f$, it holds that
\begin{align}
    f(\vec{\mu}(\mathcal{U})) = \sum_{j_1,j_2,\dots,j_k}\alpha_{j_1}\alpha_{j_2}\dots\alpha_{j_k} f(\vec{\mu}(\mathcal{C}_{j_1,\dots,j_k}))
\end{align}
and
\begin{align}
    \tr(\mathcal{U}(\rho_0)X) = \sum_{j_1,j_2,\dots,j_k}\alpha_{j_1}\alpha_{j_2}\dots\alpha_{j_k} \tr(\mathcal{C}_{j_1,\dots,j_k}(\rho_0)X).
\end{align}
Therefore, suppose it holds for all Clifford unitaries $\mathcal{C}$ that the loss of our correction is
\begin{align}
    \lvert\delta(\mathcal{C})\rvert := \lvert f(\vec{\mu}(\mathcal{C})) - \tr(\mathcal{C}(\rho_0)X) \rvert = 0.
\end{align}
Then, since $\mathcal{C}_{j_1,\dots,j_k}$ is Clifford for all $j_1,\dots,j_k$,
\begin{align}
    \lvert\delta(\mathcal{U})\rvert = \lvert &f(\vec{\mu}(\mathcal{U})) - \tr(\mathcal{U}(\rho_0)X) \rvert 
    \nonumber\\&= \left\lvert \sum_{j_1,\dots,j_k}\alpha_{j_1}\dots\alpha_{j_k} \delta(\mathcal{C}_{j_1,\dots,j_k}) \right\rvert\nonumber\\
    &= 0.
\end{align}

In other words, we have achieved zero loss on all arbitrary circuits if we obtain zero loss on training data comprised of all possible Clifford circuits. We note the a similar remark is made in  Ref.~\cite{strikis2020learning} to argue that Clifford circuits suffice for their learning-based approach to quasi-probability representation (QPR) error mitigation. However, unlike in Ref.~\cite{strikis2020learning}, depending on the channels and initial state involved in the error mitigation, there may not exist a set of parameters which achieves zero loss on all Clifford circuits. Hence, the considerations in this section should serve as high-level motivation for the linear form of the ansatz we employ, rather than a rigorous demonstration of the practicality of Clifford-based training sets.

\section{The QAOA ansatz decomposition to a quantum circuit}
\label{app:QAOA_circuit}
\begin{figure}[t]
   \includegraphics[width=0.48\textwidth]{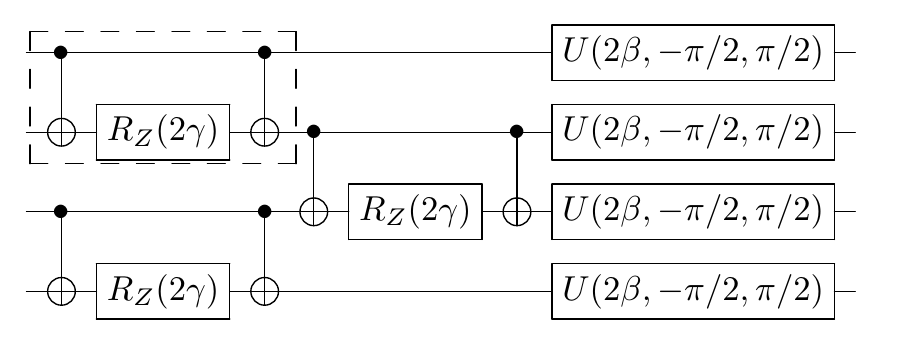}
    \caption{A layer of the  QAOA ansatz  for a 4 qubit system decomposed to IBMQ natively supported gates. Gates within the dashed contour represent  $e^{-i\gamma \sigma_Z^1 \sigma_Z^{2}}$ \cite{LANLqalgs}, while  a $U$ gate represents  $e^{-i\beta \sigma_X}$. To perform the mitigation we decompose the $U$ gates as explained in  Fig.~\ref{fig:RQC_example}. }
    \label{fig:QAOA_decompose}
\end{figure}

In Fig.~\ref{fig:QAOA_decompose} we show a decomposition of the QAOA ansatz (\ref{QAOA}) into a circuit which was used to perform the  simulations in Section~\ref{sec:results_qaoa}.

\section{Causal cone}
\label{app:causal_cone}
\begin{figure}[t]
\includegraphics[width=0.35\textwidth]{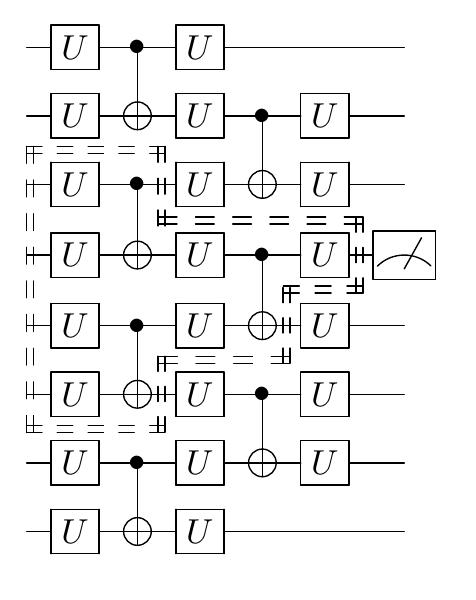}
    \caption{A causal cone of a single qubit observable is shown as gates within the dashed contour. Here we consider the case of the  hardware efficient ansatz with $Q=8$ and $p=2$.   Note that the causal cone is shown for a noisy expectation value while assuming that the Kraus matrices of a noise channel associated with a single or double qubit gate act on the same qubits as the gate. This assumption is true for our noise model.  The causal cone for a noisy circuit implementation contains the causal cone of the corresponding exact expectation value.}
    \label{fig:causal_cone_example}
\end{figure}

We define the causal cone of an observable of interest as the set of gates which affects its expectation value \cite{VidalMERA}. In the case of the shallow quantum circuits and
one or two qubit observables considered in Section~\ref{sec:results_rqc}  the causal cones contain only a fraction of all gates in the circuit. Therefore, to correct noisy expectation values of these observables it is beneficial to leave non-Clifford gates only within the causal cone, as we have done when constructing the CDR and vnCDR training sets while correcting observables from random quantum circuits. We show an example of the causal cone construction in Fig.~\ref{fig:causal_cone_example}.   

\section{Matrix Product  Operator simulation of noisy states}
\label{app:MPO}
Many-body quantum states can be represented in terms of interconnected tensors called tensor networks. Tensor networks are  basis for many standard numerical and analytical  techniques in condensed matter theory \cite{orus2019}. Matrix Product Operator (MPO) is an example of a tensor network, which corresponds to a 1-D array of tensors. In general they can describe any mixed state in the 1-D many body Hilbert space, with the dimensionality of the tensors scaling exponentially with the system size. However, states with sufficiently small  entanglement can be efficiently  represented as MPO making them a convenient numerical tool.  For a detailed introduction of MPO methods we refer the reader to Ref.~\cite{schollwock}. 

Consider an $Q$ qubit density matrix,
\begin{equation}
    \hat{\rho} = \sum_{\substack{i_{1},\dots,i_{Q}=0\\ i'_{1},\dots,i'_{Q}=0}}^{1}c_{(i_{1}\dots i_{Q})(i'_{1}\dots i'_{Q})}\ket{i_{1},\dots,i_{Q}}\bra{i'_{1},\dots,i'_{Q}}.
\end{equation}
We can express $c_{(i_{1}\dots i_{Q})(i'_{1}\dots i'_{Q})}$ as a product of matrices $W^{(1)i_{1}i'_{1}}, W^{(2)i_{2}i'_{2}}, \dots, W^{(Q)i_{Q}i'_{Q}}$, 
\begin{align}
    &c_{(i_{1}\dots i_{Q})(i'_{1}\dots i'_{Q})} = \nonumber\\ & \sum_{b_1,\dots,b_{Q-1} = 0}^{\chi-1}W^{(1)i_{1}i'_{1}}_{1,b_1}W^{(2)i_{2}i'_{2}}_{b_1,b_2}\dots W^{(Q)i_{Q}i'_{Q}}_{b_Q-1},
\end{align}
where $\chi$ is the bond dimension and $b_{1},\dots,b_{Q-1}$ are bond indices which characterize the entanglement in the state. Above for simplicity we assume the same $\chi$ for each bond index but in principle $\chi$ can be different for each of them.      For a general quantum state represented in this form one needs to use $\chi =  O(2^{n})$ \cite{schollwock}. Dropping the summation over the dummy indices we can write $\hat{\rho}$ as a MPO
\begin{align}
    \hat{\rho} = \sum_{\substack{i_{1},\dots,i_{Q}\\ i'_{1},\dots,i'_{Q}}} &W^{(1)i_{1}i'_{1}}W^{(2)i_{2}i'_{2}}\dots  \nonumber \\
    &\dots W^{(Q)i_{Q}i'_{Q}}
    \ket{i_{1},\dots,i_{Q}}\bra{i'_{1},\dots,i'_{Q}},
\end{align}

In general this is not an efficient description of a quantum state due to the exponential scaling of the bond dimension. However, for a restricted set of states an efficient representation exists, namely for states with sufficiently small entanglement. Therefore, for this restricted class of states expectation values can be classically evaluated. To be more specific let's consider $\langle X^{(1)}_1 X^{(2)}_2 \dots X^{(Q)}_Q\rangle$ where $X^{(1)}_1, X^{(2)}_2, \dots  X^{(Q)}_Q$ are single qubit observables acting respectively at  qubits $1,2,\dots,Q$.
Then 
\begin{equation}
\langle X^{(1)}_1 X^{(2)}_2 \dots X^{(Q)}_Q\rangle = Y^{(1)} Y^{(2)} \dots Y^{(Q)}, 
\end{equation}
where $Y^{(1)}, Y^{(2)}, \dots, Y^{(Q)}$ are matrices obtained as 
\begin{equation}
Y^{(1)}_1 = \sum_{i_1,i_1'} W^{(1)i_{1}i'_{1}} X^{(1)i'_1 i_1}_1, \dots
\end{equation}
Therefore, the computational cost scales as $O(\chi^2)$ enabling classical computation for small enough $\chi$.

In the case of our noise model initial noisy state corresponding to the exact state  $\hat{\rho} = |00\dots 0\rangle   \langle 00\dots 0|$ can be written as MPO with $\chi=1$ using standard MP0 techniques \cite{schollwock}. Applying these techniques one can verify that action of single qubit gates do not increase $\chi$, while action of two-qubit noisy CNOTs increase $\chi$ for bond indices linking two  qubits at which the CNOT acts  by at most factor of 16. Therefore in the case of our RQC simulation maximal $\chi$ is bounded from above by $16^{p/2}$. In practice we find that MPO representation can be further compressed after each CNOT action using standard MPO compression techniques~\cite{schollwock} to discard $W$ elements of order of  numerical precision.  
 
\section{Resource scaling}
\label{app:resource_scaling}

In the main text we systematically benchmark performance of ZNE, CDR, vnCDR error mitigation methods in the limit of an infinite number of shots. In the case of real-world quantum devices one is limited to a finite shot number. Here we analyze performance of the method in this case. We perform error mitigation of   $\langle\sigma^{1}_{X}\rangle$, $\langle\sigma^{Q/2}_{X}\rangle$,  $\langle\sigma^{1}_{Z}\sigma^{2}_{Z}\rangle$, $\langle\sigma^{Q/2}_{Z}\sigma^{Q/2+1}_{Z}\rangle$ for RQC  using $N_s=10^3,10^4,10^5$ shots per circuit and compare the results obtained with those evaluated with an infinite number of shots. To enable efficient classical simulation we consider here independent shots for each observable.  

\begin{figure*}[t]
    \centering
    \includegraphics[width=0.4\textwidth]{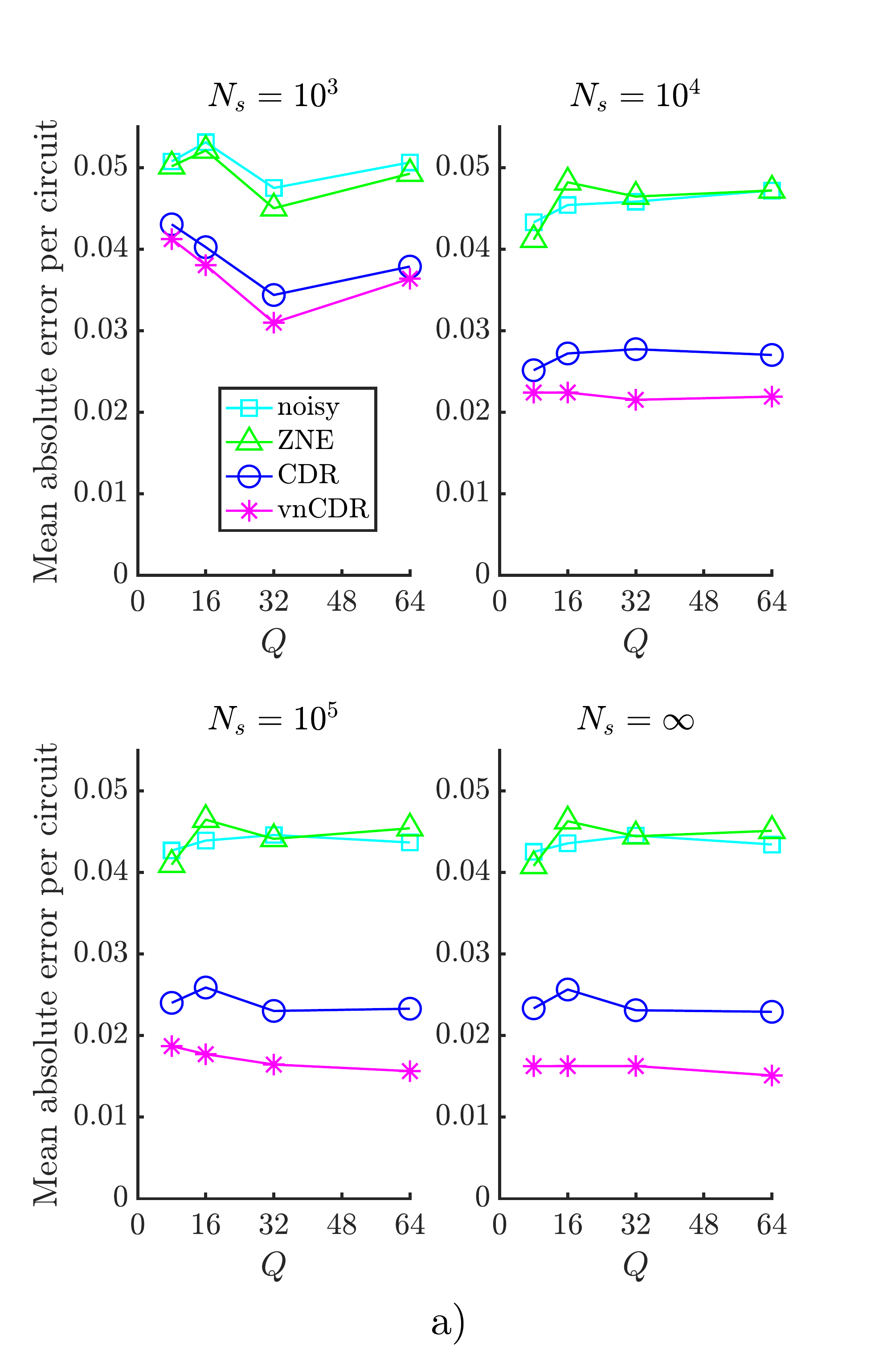}
    \includegraphics[width=0.4\textwidth]{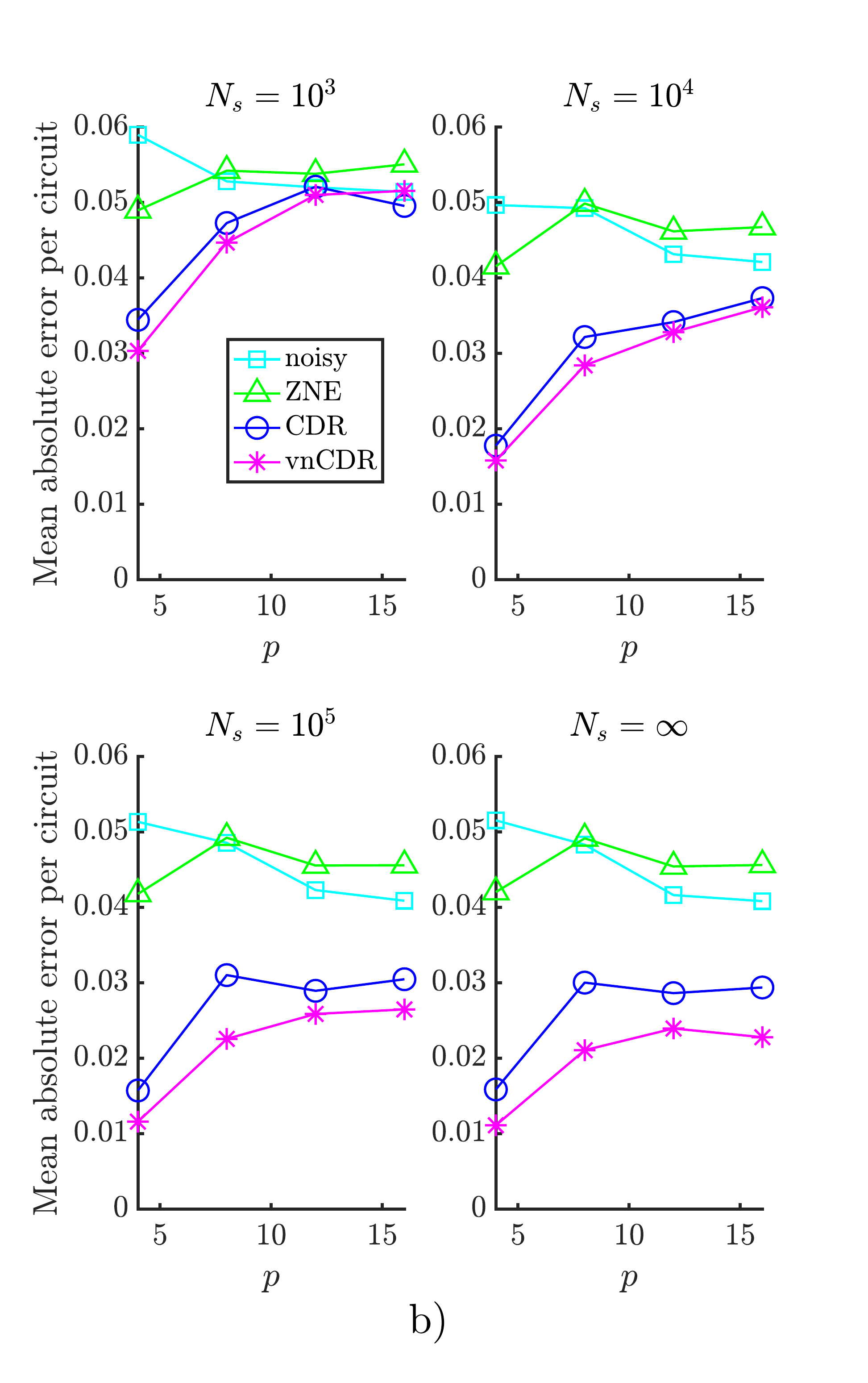}
    \caption{Correcting the IBMQ hardware efficient ansatz with random parameters various numbers of shots $N_{s}$ used to evaluate every circuit involved in each mitigation method. 
     For each value of $Q$ and $p$ we analyze 30 random circuits and  correct $\langle\sigma^{1}_{X}\rangle$, $\langle\sigma^{Q/2}_{X}\rangle$,  $\langle\sigma^{1}_{Z}\sigma^{2}_{Z}\rangle$, $\langle\sigma^{Q/2}_{Z}\sigma^{Q/2+1}_{Z}\rangle$
    for each of them.  The circuits are the same as those analyzed in Fig.~\ref{fig:RQC_scaling}. The results from  Fig.~\ref{fig:RQC_scaling} correspond to $N_s=\infty$ limit and are shown alongside finite $N_s$ results. 
    In (a) we  display the scaling with $Q$ of the mitigated and unmitigated absolute error per circuit  for $p=6$. In the left panel we show the mean values  (the solid lines), while the right panel shows the maximal values (the dashed lines).  
     In (b) the scaling with increasing $p$ for $Q=8$. 
     vnCDR method systematically outperforms ZNE and CDR methods for $N_s\ge10^4$ in the case of the shallow circuits and $N_s\ge10^5$ in the case of the deep circuits making it the method of choice for obtaining high accuracy results.   }
    \label{fig:resource_scaling}
\end{figure*}

We gather the results in Fig.~\ref{fig:resource_scaling} showing  scaling of the error  with  system size $Q$ and circuit depth $p$ for different shot costs.
We find that improvement of vnCDR over ZNE and CDR grows systematically with increasing $N_s$. We see that small $N_s=10^3$ is enough to see a systematic improvement of vnCDR over ZNE for the shallow circuits. With $N_s=10^4$ we see a systematic improvement of vnCDR over CDR for the shallow circuits and a systematic improvement of vnCDR over ZNE for the deep circuits. With $N_s=10^5$ shots we also obtain a  systematic improvement of vnCDR over CDR for the deep circuits.  We note that for setups in which vnCDR does not outperform other methods it gives results of a similar quality. We observe that increasing $N_s$ improves performance of the methods only for sufficiently small $N_s$. In the case of ZNE the results obtained with $N_s=10^3$ are of similar quality as the ones obtained in the limit of $N_s=\infty$.  For vnCDR and CDR $N_s=10^5$ is needed to that end.    

To give a full picture of the number of shots needed for the different methods one also needs to consider the number of circuits required to mitigate the circuit of interest.  ZNE requires only the execution of the circuit of interest at various noise levels. Assuming $n$ noise levels are needed with $N_{s}$ shots per circuit the total shot cost for ZNE is $n \times N_{s}$. Both CDR and vnCDR require the execution of near-Clifford training circuits as well as the circuit of interest. Assuming a training set consisting of $m$ circuits, each run using $N_{s}$ shots the total shot cost for CDR is given as $(m+1) \times N_{s}$. vnCDR requires the Clifford training circuits and the circuit of interest implemented at various noise levels. With $n$ noise levels and $m$ training circuits each evaluated using $N_{s}$ shots the total shot cost for vnCDR is $(m+1) \times n \times N_{s}$. For our RQC results $n = 5$ and $m = 100$. Therefore, the shot cost for ZNE is given as $5 \times N_{s}$, for CDR $101 \times N_{s}$ and for vnCDR is $501 \times N_{s}$. Then to see systematic improvement over ZNE and CDR,  we need respectively $5\times10^5-5\times10^6$  and $5\times10^6-5\times10^7$ shots in total. These shot numbers can be obtained with current devices proving usefulness of vnCDR.

\end{document}